\documentclass[12pt,a4paper]{article}
\usepackage{latexsym,amssymb,amsmath,amsthm}

\newtheorem{theorem}{Theorem}

\begin{document}
\title{Implementation of linear maps with circulant matrices via modulo 2 rectifier circuits of bounded depth\footnote{Research supported in part by RFBR, grants
11--01--00508, 11--01--00792, and OMN RAS ``Algebraic and
combinatorial methods of mathematical cybernetics and information
systems of new generation'' program (project ``Problems of optimal
synthesis of control systems'').}}
\date{}
\author{Igor S. Sergeev\footnote{e-mail: isserg@gmail.com}}

\maketitle

\begin{abstract}
In the present note we show that for any constant $k \in \mathbb
N$ an arbitrary Boolean circulant matrix can be implemented via
modulo~2 rectifier circuit of depth $2k-1$ and complexity
$O\left(n^{1+1/k}\right)$, and also via circuit of depth $2k$ and
complexity $O\left(n^{1+1/k}\log^{-1/k} n\right)$.
\end{abstract}

Recall that {\it rectifier $(m,n)$-circuit} is an oriented graph
with $n$ vertices labeled as inputs and $m$ vertices labeled as
outputs. {\it Modulo $2$ rectifier circuit} implements a Boolean
$m\times n$ matrix $A=(a_{i,j})$ iff for any $i$ and $j$ the
number of oriented paths from $j$-th input to $i$-th output is
congruent to $a_{i,j}$ modulo~2. Complexity of a circuit is the
number of edges in it, circuit depth is the maximal length of an
oriented path. See details in~\cite{jue,lue}.

$n\times n$ matrix $Z=(z_{i,j})$ is {\it circulant} iff for any
$i$, $j$ one has $z_{i,j} = z_{0,k}$, where $k=(j-i)\bmod n$.

Consider a linear map with Boolean circulant $n\times n$
matrix~--- it computes a cyclic (algebraic, over $GF(2)$)
convolution with some constant vector $A$. Indeed, components of
vector $C=(C_0,\ldots,C_{n-1})$ which is a convolution of vectors
$A=(A_0,\ldots,A_{n-1})$ and $B=(B_0,\ldots,B_{n-1})$ satisfy
formulae:
$$ C_k = \sum_{i+j \equiv k \mod n} A_iB_j. $$

The following theorem allows to extend results~\cite{gse,gs1e} on
comparison of complexity of implementation of some circulant
matrices via rectifier circuits and modulo 2 rectifier circuits to
bounded depth circuits.

\begin{theorem}
For any $k \in \mathbb N$ an arbitrary Boolean circulant $n\times
n$ matrix $Z$ can be implemented via modulo $2$ rectifier circuit:

a) of depth $2k-1$ and complexity at most $f(2k-1)n^{1+1/k}$;

b) of depth $2k$ and complexity at most $f(2k)n\left(\frac{n}{\log
n}\right)^{1/k}$.
\end{theorem}

The proof is by induction. For $k=1$ we use a trivial depth-1
circuit of complexity $O(n^2)$ and a circuit of depth 2 and
complexity $O(n^2/\log n)$ provided by O.B. Lupanov's
method~\cite{lue}.

Now we prove an induction step from $k-1$ to $k$. We use the
polynomial multiplication method due to A.L. Toom~\cite{toe}
together with À.~Sch\"onhage's idea~\cite{she} allowing to extend
the method to binary polynomials. The depth-$d$ polynomial
multiplication is reduced to several parallel depth-$(d-2)$
multiplications.

Split a vector of variables into $q$ blocks of length $n/q$ and
interpret each block as a vector of coefficients of a polynomial
from the ring
$$ R=GF(2)[y]/(y^{2\cdot3^s}+y^{3^s}+1).$$
Parameter $s$ satisfies condition $3^s \ge n/q$.

So, multiplication of binary polynomials of degree $n-1$ can be
performed as a multiplication of polynomials of degree $n/q-1$
over $R$. The latter multiplication can be performed via DFT of
order $3^m \ge 2q$ with primitive root $\zeta=y^{s+1-m}\in R$.

Next, we describe a circuit.

Its input is a polynomial $B(x)=\sum B_i x^i \in R[x]$ of degree
$q-1$. A constant factor is denoted by $A(x)=\sum A_i x^i$. Output
is the product $C(x)=A(x)B(x)=\sum C_i x^i$.

1. Compute $B(\zeta^0)$, \ldots, $B(\zeta^{3^m-1})$.

2. Compute $C(\zeta^i)=A(\zeta^i)B(\zeta^i)$ for all
$i=0,\ldots,3^m-1$.

3. Compute coefficients of $C(x)$.

We implement stages 1 and 3 via depth-1 circuits and stage 2 ---
via circuit of depth $d-2$. Next, we estimate the circuit
complexity, denote it by $M(d,n)$.

1. Multiplication by a power of $y$ in $R$ has linear complexity.
Hence, the value of polynomial $F(x)$ at the point $y^p$ can be
computed with linear complexity as well. Therefore, the complexity
of stage 1 is $O(3^m3^sq)$.

2. Every multiplication at the stage 2 is a multiplication of
binary polynomials of degree $2\cdot3^s-1$ with a subsequent
modulo reduction. Perform multiplication via circuit of depth
$d-2$ and complexity $M(d-2,2\cdot3^s)$ provided by induction
hypothesis. Reduction of a polynomial $g(y)$ (here, of degree
$4\cdot3^s-1$) modulo $y^{2\cdot3^s}+y^{3^s}+1$ is performed via
duplication of some of its coefficients (that is, outputs of a
preceding subcircuit) and identifying of some coefficients, since
every coefficient of $g$ is to be used at most twice. Henceforth,
modulo reduction can be embedded into multiplication circuit with
no depth increasing and with at most doubling of the circuit
complexity. Thus, the total complexity of stage 2 is at most
$2\cdot3^mM(d-2,2\cdot3^s)$.

3. By the fundamental property of DFT, coefficients of $C(x)$
satisfy $C_i=C^*(\zeta^{-i})$, where polynomial $C^*(x)$ has
coefficients $C(\zeta^i)$. Thus, the complexity of stage 3 is that
of stage 1, $O(3^m3^sq)$.

To transform the product of polynomials over $R$ backward to the
product of binary polynomials, one performs a substitution
$x=y^{2\cdot3^s}$. The substitution preserves depth and complexity
of the circuit.

The choice of parameters to obtain required complexity bounds is:
$q=n^{1/k}$ for $d=2k-1$ and $q=(n/\log n)^{1/k}$ for $d=2k$;
$3^s=\Theta(n/q)$, $3^m=\Theta(q)$.

(By construction, $f(k)=O(c^k)$ for some constant $c$.) \qed

\end{document}


\title{Реализация линейных преобразований с циркулянтными матрицами вентильными схемами по модулю 2
ограниченной глубины\footnote{Работа выполнена при финансовой
поддержке РФФИ, проекты 11--01--00508 и 11--01--00792, и программы
фундаментальных исследований ОМН РАН <<Алгебраические и
комбинаторные методы математической кибернетики и информационные
системы нового поколения>> (проект <<Задачи оптимального синтеза
управляющих систем>>).}}
\date{}
\author{И. С. Сергеев\footnote{эл. адрес: isserg@gmail.com}}

\maketitle

\begin{abstract}
В настоящей заметке показано, что при любой постоянной $k \in
\mathbb N$ произвольную булеву циркулянтную матрицу можно
реализовать вентильной схемой по модулю 2 глубины $2k-1$ и
сложности $O\left(n^{1+1/k}\right)$, а также схемой глубины $2k$ и
сложности $O\left(n^{1+1/k}\log^{-1/k} n\right)$.
\end{abstract}

Напомним, что вентильная $(m,n)$-схема~--- это ориентированный
ациклический граф, в котором $n$ вершин отмечены как входы и $m$
вершин отмечены как выходы. Вентильная схема по модулю 2 реализует
булеву $m\times n$-матрицу $A=(a_{i,j})$ тогда и только тогда,
когда при любых $i$ и $j$ число ориентированных путей из $j$-го
входа в $i$-й выход сравнимо с $a_{i,j}$ по модулю 2. Сложностью
схемы называется число ребер в ней, а глубиной~--- максимальная
длина ориентированного пути. Подробнее см. в~\cite{lu,ju}.

$n\times n$-матрица $Z=(z_{i,j})$ называется циркулянтной, если
при любых $i$, $j$ выполнено $z_{i,j} = z_{0,k}$, где
$k=(j-i)\bmod n$.

Рассмотрим линейный булев оператор размера $n\times n$ с
циркулянтной матрицей --- он реализует циклическую
(алгебраическую, над $GF(2)$) свертку с некоторым постоянным
вектором $A$. Действительно, компоненты вектора
$C=(C_0,\ldots,C_{n-1})$~--- свертки векторов
$A=(A_0,\ldots,A_{n-1})$ и $B=(B_0,\ldots,B_{n-1})$, выражаются
формулами:
$$ C_k = \sum_{i+j \equiv k \mod n} A_iB_j. $$

Следующая теорема позволяет распространить
результаты~\cite{gs,gs1} о сравнении сложности реализации
некоторых циркулянтных матриц вентильными схемами и вентильными
схемами по модулю 2 на схемы ограниченной глубины.

\begin{theorem}
При $k \in \mathbb N$ произвольную булеву циркулянтную $n\times
n$-матрицу $Z$ можно реализовать вентильной схемой по модулю $2$:

а) глубины $2k-1$ и сложности не более $f(2k-1)n^{1+1/k}$;

б) глубины $2k$ и сложности не более $f(2k)n\left(\frac{n}{\log
n}\right)^{1/k}$.
\end{theorem}

Рассуждение проведем по индукции. При $k=1$ предъявляются
тривиальная схема глубины 1 и сложности $O(n^2)$, а также схема
глубины 2 и сложности $O(n^2/\log n)$, которая строится методом
О.Б.~Лупанова~\cite{lu}.

Докажем индуктивный переход от $k-1$ к $k$. Для доказательства
используется метод А.Л.~Тоома умножения многочленов~\cite{to}
вместе с приемом А.~Ш\"енхаге~\cite{sh}, позволяющим
распространить метод на двоичные многочлены. Умножение многочленов
с глубиной $d$ сводится к нескольким параллельным умножениям
глубины $d-2$.

Разобьем вектор переменных на $q$ частей длины $n/q$. Эти части
интерпретируем как вектора коэффициентов многочленов из кольца
$$ R=GF(2)[y]/(y^{2\cdot3^s}+y^{3^s}+1)$$ при младших степенях $y$.
Параметр $s$ выбирается из условия $3^s \ge n/q$.

Сведем умножение двоичных многочленов степени $n-1$ к умножению
многочленов степени $n/q-1$ над $R$. Последнее умножение выполним
при помощи ДПФ порядка $3^m \ge 2q$ с примитивным корнем
$\zeta=y^{s+1-m} \in R$.

Опишем схему.

На входе многочлен $B(x)=\sum B_i x^i \in R[x]$ степени $q-1$.
Постоянный сомножитель обозначим через $A(x)=\sum A_i x^i$. На
выходе --- произведение $C(x)=A(x)B(x)=\sum C_i x^i$.

1. Вычисляем $B(\zeta^0)$, \ldots, $B(\zeta^{3^m-1})$.

2. Вычисляем $C(\zeta^i)=A(\zeta^i)B(\zeta^i)$ для всех
$i=0,\ldots,3^m-1$.

3. Вычисляем коэффициенты многочлена $C(x)$.

Этапы 1 и 3 реализуем схемами глубины 1, а этап 2 --- схемой
глубины $d-2$. Оценим сложность схемы, обозначим ее $M(d,n)$.

1. Умножение на степень $y$ в кольце $R$ выполняется с линейной
сложностью, поэтому с линейной сложностью вычисляется значение
многочлена $F(x)$ в точке $y^p$. Следовательно, сложность первой
схемы оценивается как $O(3^m3^sq)$.

2. Каждое из умножений на шаге 2 есть умножение двоичных
многочленов степени $2\cdot3^s-1$ с последующим приведением по
модулю. Умножение выполним методом из индуктивного предположения
глубины $d-2$ и сложности $M(d-2,2\cdot3^s)$. Приведение
многочлена (в данном случае, степени $4\cdot3^s-1$) по модулю
$y^{2\cdot3^s}+y^{3^s}+1$ достигается дублированием некоторых его
коэффициентов (выходов предшествующей схемы) и отождествлением
некоторых коэффициентов, т.к. каждый коэффициент приводимого
многочлена используется не более чем дважды. Следовательно,
приведение по модулю можно реализовать без увеличения глубины и с
не более чем двукратным увеличением сложности. Общую сложность
второго этапа теперь можно оценить как
$2\cdot3^mM(d-2,2\cdot3^s)$.

3. Согласно основному свойству ДПФ искомые коэффициенты многочлена
$C(x)$ находятся как $C_i=C^*(\zeta^{-i})$, где многочлен $C^*(x)$
имеет коэффициенты $C(\zeta^i)$. Поэтому сложность этапа 3 можно
оценить так же, как и сложность этапа 1, $O(3^m3^sq)$.

Чтобы результат умножения многочленов над $R$ преобразовать
(обратно) в результат умножения двоичных многочленов, надо
выполнить подстановку $x=y^{2\cdot3^s}$ и привести подобные. Это
преобразование реализуется отождествлением выходов и не влияет на
глубину и сложность схемы.

Оценки теоремы получаются при следующем выборе параметров:
$q=n^{1/k}$ при $d=2k-1$ и $q=(n/\log n)^{1/k}$ при $d=2k$;
$3^s=\Theta(n/q)$, $3^m=\Theta(q)$.

(По построению, $f(k)=O(c^k)$ при некоторой константе $c$.) \qed